\documentclass[aps, prl, superscriptaddress,reprint]{revtex4-1}
\usepackage{amssymb}
\usepackage{hyperref}
\usepackage{amsmath}
\usepackage{graphicx}
\usepackage{bm}
\usepackage[export]{adjustbox}
\usepackage{overpic}
\usepackage{xcolor}

\begin{document}
\title{Enhanced Superexchange in a Tilted Mott Insulator}

\author{Ivana Dimitrova}
\affiliation{Research Laboratory of Electronics, MIT-Harvard Center for Ultracold Atoms,
Department of Physics, Massachusetts Institute of Technology, Cambridge, Massachusetts 02139, USA}
 \author{Niklas Jepsen}
  \affiliation{Research Laboratory of Electronics, MIT-Harvard Center for Ultracold Atoms,
Department of Physics, Massachusetts Institute of Technology, Cambridge, Massachusetts 02139, USA}
 \author{Anton Buyskikh}
  \affiliation{Department of Physics and SUPA, University of Strathclyde, Glasgow G4 0NG, United Kingdom}
 \author{Araceli Venegas-Gomez}
  \affiliation{Department of Physics and SUPA, University of Strathclyde, Glasgow G4 0NG, United Kingdom}
\author{Jesse Amato-Grill}
\affiliation{Research Laboratory of Electronics, MIT-Harvard Center for Ultracold Atoms,
Department of Physics, Massachusetts Institute of Technology, Cambridge, Massachusetts 02139, USA}
 \affiliation{Department of Physics, Harvard University, Cambridge, Massachusetts 02138, USA}
 \author{Andrew Daley}
\affiliation{Department of Physics and SUPA, University of Strathclyde, Glasgow G4 0NG, United Kingdom}
 \author{Wolfgang Ketterle}
  \affiliation{Research Laboratory of Electronics, MIT-Harvard Center for Ultracold Atoms,
Department of Physics, Massachusetts Institute of Technology, Cambridge, Massachusetts 02139, USA}

\date{\today}

\begin{abstract}
In an optical lattice entropy and mass transport by first-order tunneling is much faster than spin transport via superexchange. Here we show that adding a constant force (tilt) suppresses first-order tunneling, but not spin transport, realizing new features for spin Hamiltonians. Suppression of the superfluid transition can stabilize larger systems with faster spin dynamics. For the first time in a many-body spin system, we vary superexchange rates by over a factor of 100 and tune spin-spin interactions via the tilt. In a tilted lattice, defects are immobile and pure spin dynamics can be studied.
\end{abstract}
\maketitle

The importance of spin systems goes far beyond quantum magnetism. Many problems in physics can be mapped onto spin systems. Famous examples are the Jordan-Wigner transformation between spin chains and lattice fermions, and the mapping of neural networks to Ising models. The study of spin Hamiltonians has provided major insights into phase transitions and non-equilibrium physics.  Therefore, the properties of well controlled spin systems are explored using various platforms \cite{georgescu14}.  

In the field of ultracold atoms, such Hamiltonians are realized by a mapping from the Hubbard model in the Mott insulating (MI) state to Heisenberg models with effective spin-spin coupling given by a second order tunneling process (superexchange) \cite{ddl03, svistunov03}. Although immense progress has been made towards the realization of spin-ordered ground states \cite{blochreview05, lewenstein12, gross17, hofstetter18}, a major challenge is to reach low spin temperatures. A promising route is adiabatic state preparation \cite{schachenmayer15}, but in a trapped system a higher entropy region surrounds a low-entropy MI core, whose ultimate temperature and lifetime is limited in most cases by mass or energy transport. A fundamental limitation of superexchange-driven schemes is that the lattice depth controls both mass transport (occuring at the tunneling rate $t/\hbar$) and the effective spin dynamics (at $t^2/(\hbar U)$, where $U$ is the on-site interaction). Schemes isolating the MI by shaping the trapping potential have been proposed \cite{chiu18, bernier09, mathy12, ho09, mazurenko17}. 

Here we use a controlled potential energy offset between neighboring sites (a tilt) to decouple spin transport from density dynamics in the MI regime. Tilted lattices have been used before to suppress tunneling (in spin-orbit coupling schemes with laser-assisted tunneling \cite{hiro13, monika13, colin15, monika15}), or to implement spin models using resonant tunneling between sites with different occupations \cite{sachdev02, greiner11, nagerl13, nagerl14}. Energy offsets have been used in double-well potentials to modify superexchange rates \cite{trotzky08}, between sublattices to suppress first-order tunneling and to observe magnetization decay via superexchange \cite{brown15}.  

\begin{figure}[t!]
\center
\includegraphics[width=0.3 \textwidth]{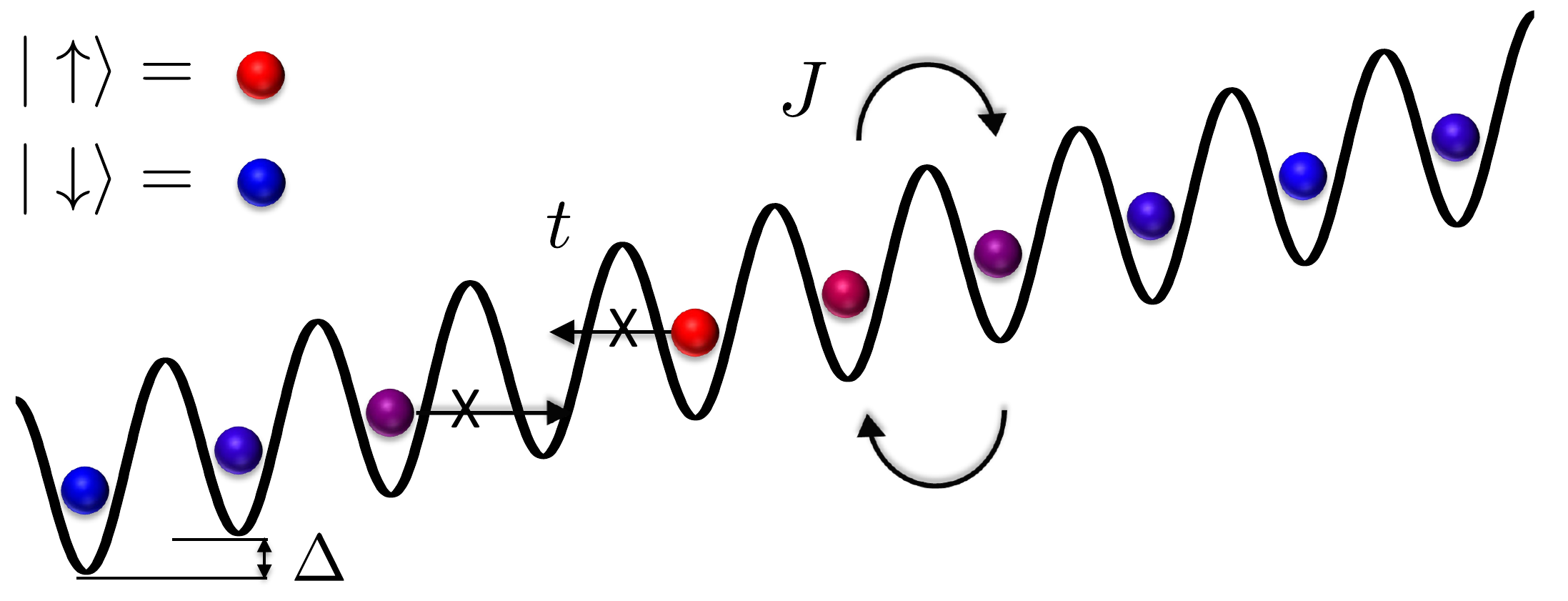}
\caption{In a tilted lattice with energy offset per site $\Delta$, tunneling at $t/\hbar$ is suppressed, while superexchange at $J(\Delta)/\hbar$ is still allowed. This enables the slower superexchange processes to dominate the dynamics even in systems with defects. }
\label{setup}
\end{figure}

The implications of using an off-resonant tilt for studying spin physics fall into four categories: (i) A tailored density distribution can be chosen which is frozen-in by the tilt. (ii) The tilt suppresses the transition to a superfluid (SF). We use these two features to stabilize larger MI plateaus at lower lattice depths. (iii) The sign and magnitude of the superexchange interaction can be tuned with the tilt which allows access to a larger range of magnetic phases. (iv) In a tilted MI with $n$ atoms per site, number defects ($n\,{\pm}\,1$) are localized. This turns $t\text{-}J$ models \cite{Lee06} into spin models with static impurities and allows the study of pure spin dynamics.

In a tilted lattice, the energy difference between lattice sites prevents first-order tunneling. More precisely, the dynamics of a single particle are Bloch oscillations \cite{dahan96, preiss15} and if the tilt per site $\Delta$ is larger than the bandwidth, their amplitude is smaller than a lattice site. In contrast, swapping particles incurs no energy cost, preserving superexchange (Fig.\,\ref{setup}), but with a modified matrix element. For $n\,{=}\,1$ it is \cite{trotzky08}:
\begin{equation}
J(\Delta)=\frac{4t^2}{U}\frac{1}{2}\left(\frac{1}{1-\Delta/U}+\frac{1}{1+\Delta/U} \right)
\label{JDelta}
\end{equation}
where tunneling resonances at $\Delta\,{=}\,U/m$ ($m\,{=}\,1,2,3...$) \cite{nagerl14} should be avoided. We implement the tilt with an AC Stark shift gradient from a far-detuned 1064$\,$nm laser beam, offset by a beam radius from the sample. We load a $^7$Li Bose-Einstein condensate \cite{ivana17} into a 3D 1064$\,$nm optical lattice in the MI regime. Although the tilt can be applied in any direction, here we use a tilt only along one axis of the lattice and study 1D dynamics (see \cite{supp}). 

\textit{(i) Preparing large non-equilibrium MI plateaus.} In most optical lattice experiments, the number of atoms (and therefore the signal-to-noise ratio of measurements) is not determined by the number of available atoms from the cooling cycle, but by the available laser power (and therefore beam size) for the optical lattice. This determines the harmonic confinement potential at each lattice depth. The equilibrium size of a MI plateau with $n$ atoms per site is determined by the balance between the local chemical potential $\mu\,{\approx}\, n\, U$ and the harmonic trapping potential. Its radius $r\,{\propto}\,\mu^{1/2}$, so the total atom number $N \propto U^{3/2}$, where $U$ is controlled by the scattering length $a$ via a Feshbach resonance \cite{jesse19}. We find that the $n\,{=}\,1$ MI plateau loaded at $a\,{=}\,300\,a_0$ has an order of magnitude more atoms than the one loaded at $a\,{=}\,50\,a_0$ (see Fig.\,S3 in \cite{supp}). We initialize the experiment by loading 45,000 atoms at $a\,{=}\,300\,a_0$ at a lattice depth $V\,{=}\,35\,E_R$ in a pure $n\,{=}\,1$ MI with diameter of 40-45 sites and then freeze in this distribution by applying a tilt with a 300$\,\mu$s linear ramp, much faster than $\hbar/t\,{=}\,28\,$ms. This allows the decoupling of MI state preparation from further spin experiments, which could be carried out at very different scattering lengths and lattice depths.  

\begin{figure}[h!]
\center
\begin{overpic}[width=0.4 \textwidth]{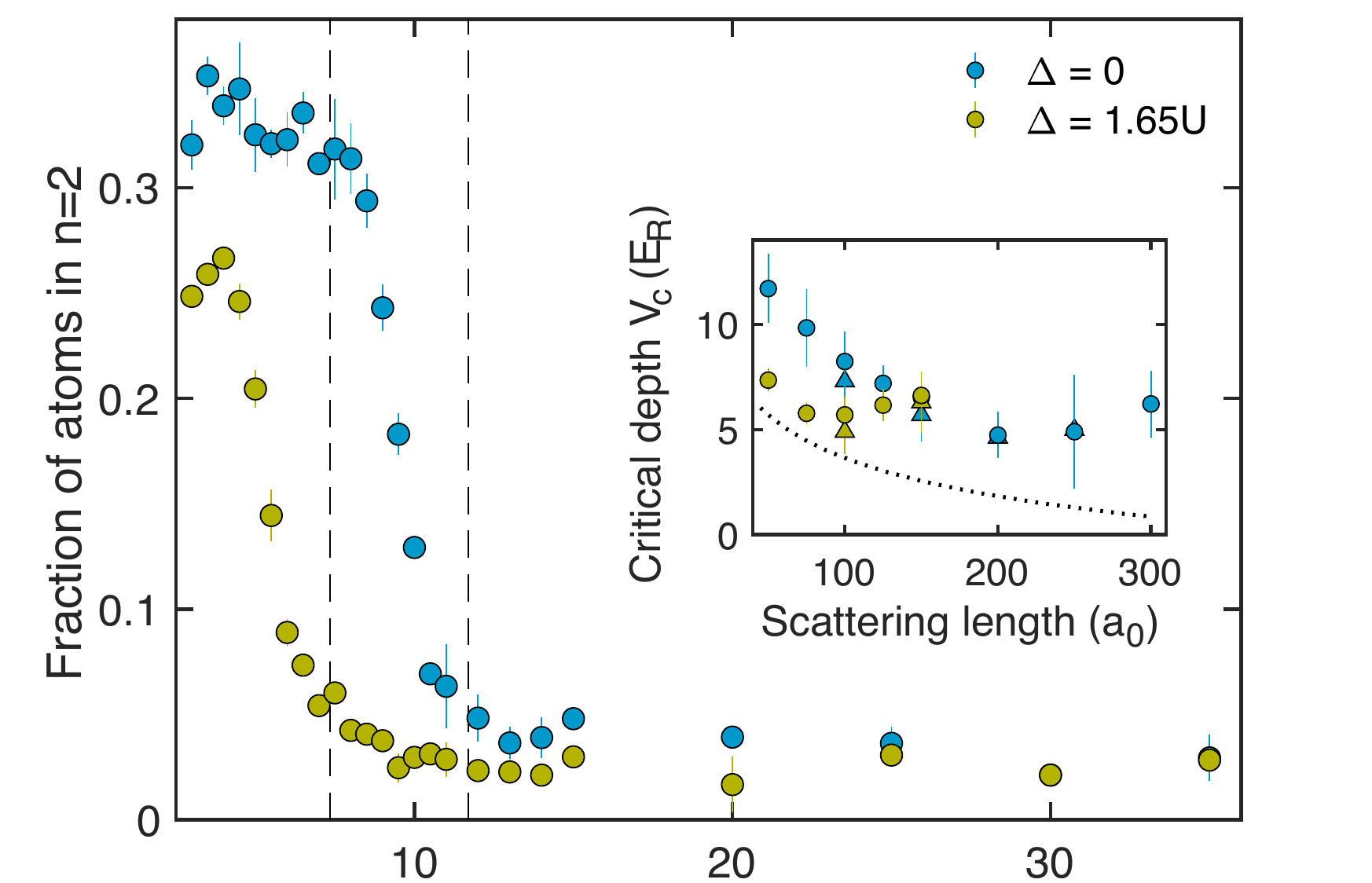}\put(-4,60){\bf{a)}}\end{overpic}
\begin{overpic}[width=0.4 \textwidth]{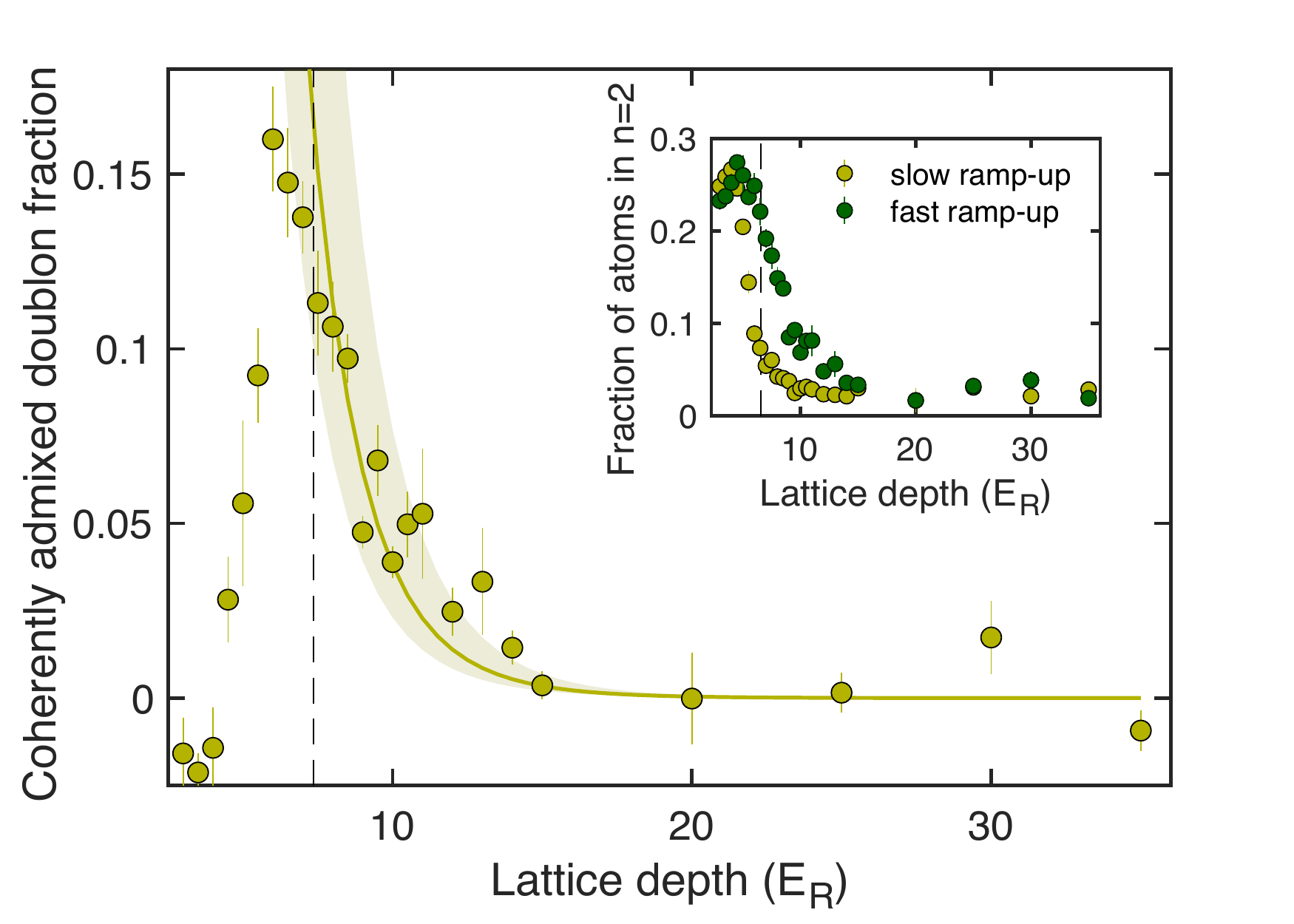}\put(-4,62){\bf{b)}}\end{overpic}
\caption{Stabilization of large Mott plateaus at small lattice depths. a) Fraction of atoms in doubly occupied sites as a function of lattice depth $V_z$ measured with (gold) and without (blue) a tilt at $a\,{=}\,50\,a_0$. Inset: Critical lattice depth $V_c$ (dashed lines in main plot) below which the fraction of atoms in $n\,{=}\,2$ is more than 3\% above the noise floor. Two initial density distributions are used: (i) nonequilibrium: an $n\,{=}\,1$ plateau prepared at $a\,{=}\,300\,a_0$ (circles) and (ii) equilibrium: an $n\,{=}\,1$ plateau prepared at the final scattering length (triangles). The dotted line corresponds to the SF-MI transition. b) Virtual and real doublon populations at $a\,{=}\,50\,a_0$ and $\Delta\,{=}\,1.65\,U$. The fraction of coherently admixed doublons is the difference between the doublons from the fast and the slow ramp, shown in the inset. Solid line: probability of doublon admixture. The shaded region accounts for tilt inhomogeneity. The dashed lines corresponds to $V_c$. The negative values of the coherent doublon fraction are an artefact in the breakdown regime.  }
\label{doublons}
\end{figure}

\textit{(ii) Increasing the speed of superexchange}.
The speed of superexchange is proportional to $t^2$ and therefore increases dramatically at lower lattice depths. Due to competing heating and loss processes, most experiments on spin physics are carried out at lattice depths only slightly above the SF-MI transition. The melting of the Mott plateaus at the transition can be suppressed by a tilt and spin Hamiltonians can be studied at lattice depths even below the phase transition. Next, we experimentally determine how much the lattice depth can be lowered.   

We associate the breakdown of the initial MI plateau with the appearance of doublons  (two atoms per site \cite{strohmaier10}), which are measured by interaction spectroscopy. We transfer only atoms in $n\,{=}\,2$ sites to another hyperfine state by using the interaction-shifted transition frequency, so that atoms in $n\,{=}\,1$ sites are not affected \cite{jesse19}.
After loading, we decrease the scattering length, lower the lattice depth $V_z$ along the direction of the tilt while keeping $V_{x}\,{=}\,V_y\,{=}\,35\,E_R$, and hold for 10$\,$ms. We detect doublons by ramping $V_z$ back to 35$\,E_R$ on a timescale ${\sim}\, \hbar/t$ but slower than $\hbar/U$, so that there is local (but not global) equilibrium.  The fraction of atoms on $n\,{=}\,2$ sites at $a\,{=}\,50\,a_0$ is shown in Fig.\,\ref{doublons}a. Below a critical lattice depth $V_c$, a sharp increase in the number of doublons is observed. Without the tilt $V_c\,{=}\,11.7\,E_R$, while with a tilt of $\Delta\,{=}\, 1.65\, U$, $V_c\,{=}\,7.3\,E_R$, implying an increase in the superexhchange rate from Eq.\,\eqref{JDelta} by a factor of 5 at the critical depth in the tilted lattice. 

To generalize this result, we repeat the measurement at several scattering lengths (inset of Fig.\,\ref{doublons}a). All $V_c$ are above the threshold for the SF-MI transition because of the spatial shrinking of the equilibrium Mott plateaus in a harmonic trap \cite{monien98, ejima11}. Without the tilt, $V_c$ is determined by the proximity to the SF-MI transition and the breakdown of plateaus is driven by first-order tunneling in the single-band approximation. Note that global density redistribution is not responsible for the breakdown in this measurement, as indicated by the fact that when the lattice is loaded at the final scattering length, so that little or no density redistribution is expected, we see similar $V_c$ (triangles in Fig.\,\ref{doublons}a). 

With the tilt, this melting can be suppressed and we observe that $V_c$ is decreased, resulting in faster spin dynamics. However, we observe that we cannot stabilize the Mott plateaus by tilts for lattice depths smaller than $V_c\,{\approx}\, 6.3\, E_R$ which we interpret as a breakdown of the single-band approximation. We find $V_c$ to have only a weak dependence on tilt for the range of tilts used (0.3 to 0.9$\,E_R$). Note that at this lattice depth, the bandgap is $3\, E_R$, and the width of the first excited band is $1.6\,E_R$. At a somewhat lower lattice depth of $4\, E_R$, we observe that atoms are accelerated out of the lattice, a clear sign for the breakdown of single-band physics. In cubic 2D and 3D lattices, the motion separates in x, y, and z and the effective breakdown of the single-band approximation should be independent of dimension. Assuming that the lattice can be lowered to $6.3\,E_R$ in 3D, then at $a\,{=}\,100\,a_0$ where the SF-MI transition is at $V_c\,{=}\,13.3\,E_R$, superexchange can be 50 times faster at $\Delta/U\,{=}\,1.4$, where $J/\hbar$ in Eq.\,\eqref{JDelta} is the same as for $\Delta=0$.  

The tilt suppresses the real population of doublons, responsible for the breakdown of MI plateaus, but not the virtual ones (coherent doublon admixtures), responsible for superexchange. In leading order in perturbation theory in $t$, the $n\,{=}\,1$ MI ground state has doublon-hole admixtures with probability $P\,{=}\,2t^2/(U\,{-}\,\Delta)^2\,{+}\,2t^2/(U\,{+}\,\Delta)^2$. These admixtures are taken into account by a unitary transformation which leads to the \textit{effective} spin Hamiltonian acting on the unperturbed states \cite{cohen-tannoudji76, svistunov03}. As perturbation theory breaks down when $t$ and $U$ become comparable, the distinction between real and virtual doublons is blurred. Virtual doublons have been detected without a tilt in \cite{bloch05, waseem10}. We measure the number of coherently admixed doublon-hole pairs as the difference between all doublons (measured with a lattice ramp-up faster than $\hbar/U$, projecting the wavefunction onto Fock states) and the real doublons (incoherent doublons, measured with a slow, locally adiabatic ramp-up as in Fig.\,\ref{doublons}a). Fig.\,\ref{doublons}b shows that the presence of the tilt does not inhibit this coherent admixture, but only modifies its probability. At $V_c$ perturbation theory breaks down.
 
\textit{(iii) Tuning the Heisenberg parameters with a tilt.} Tilts comparable to $U$ tune the strength and sign of the superexchange interactions (Eq.\,\eqref{JDelta}). This effect has so far only been observed for two particles in a double-well \cite{trotzky08}. Here we demonstrate it for the first time in a many-body system by measuring the relaxation dynamics of a nonequilibrium state in a spin chain.

\begin{figure}[t!]
\center
\begin{overpic}[width=0.4 \textwidth]{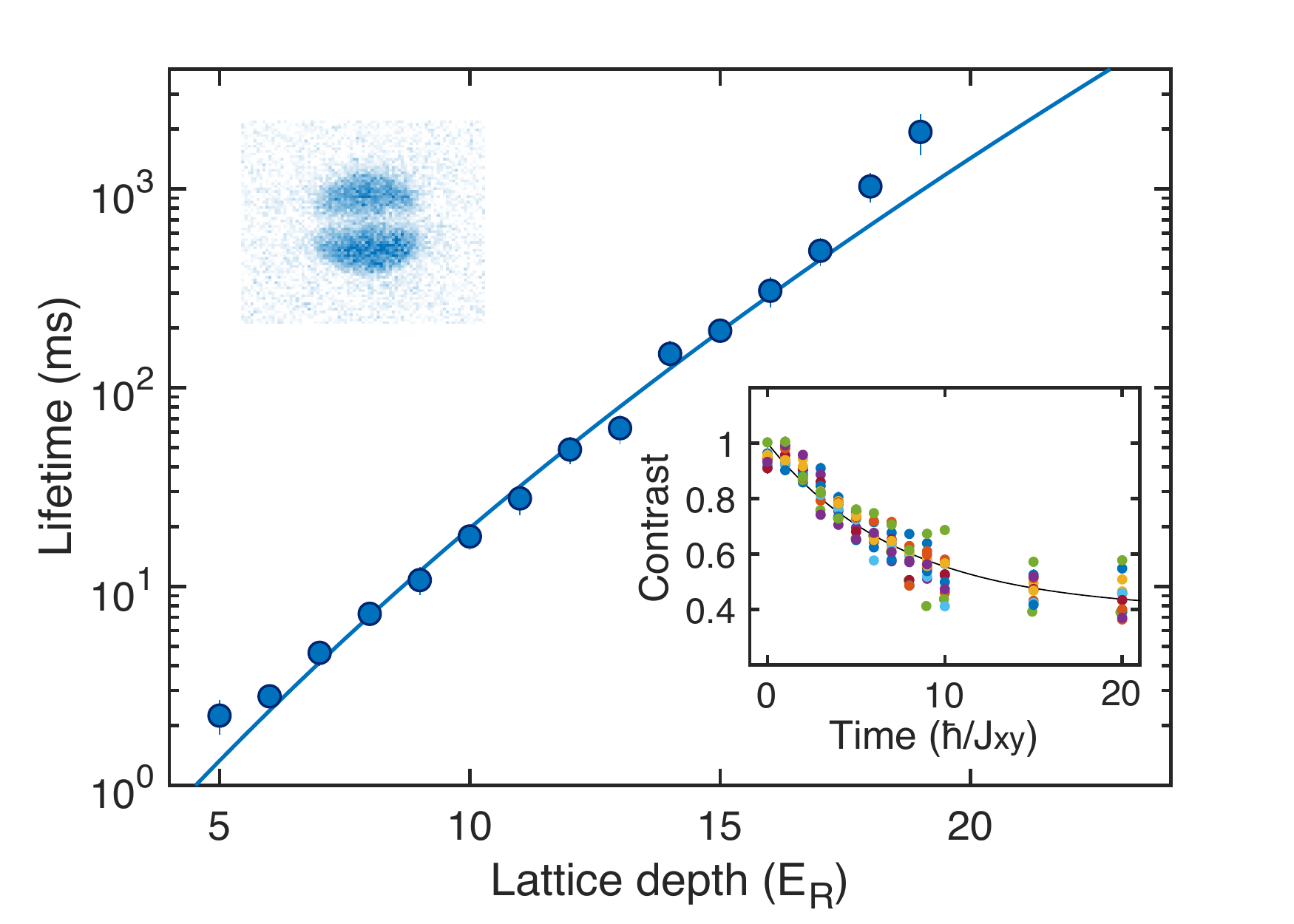}\put(0,62.5){\bf{a)}}
\end{overpic}
\begin{overpic}[width=0.4 \textwidth]{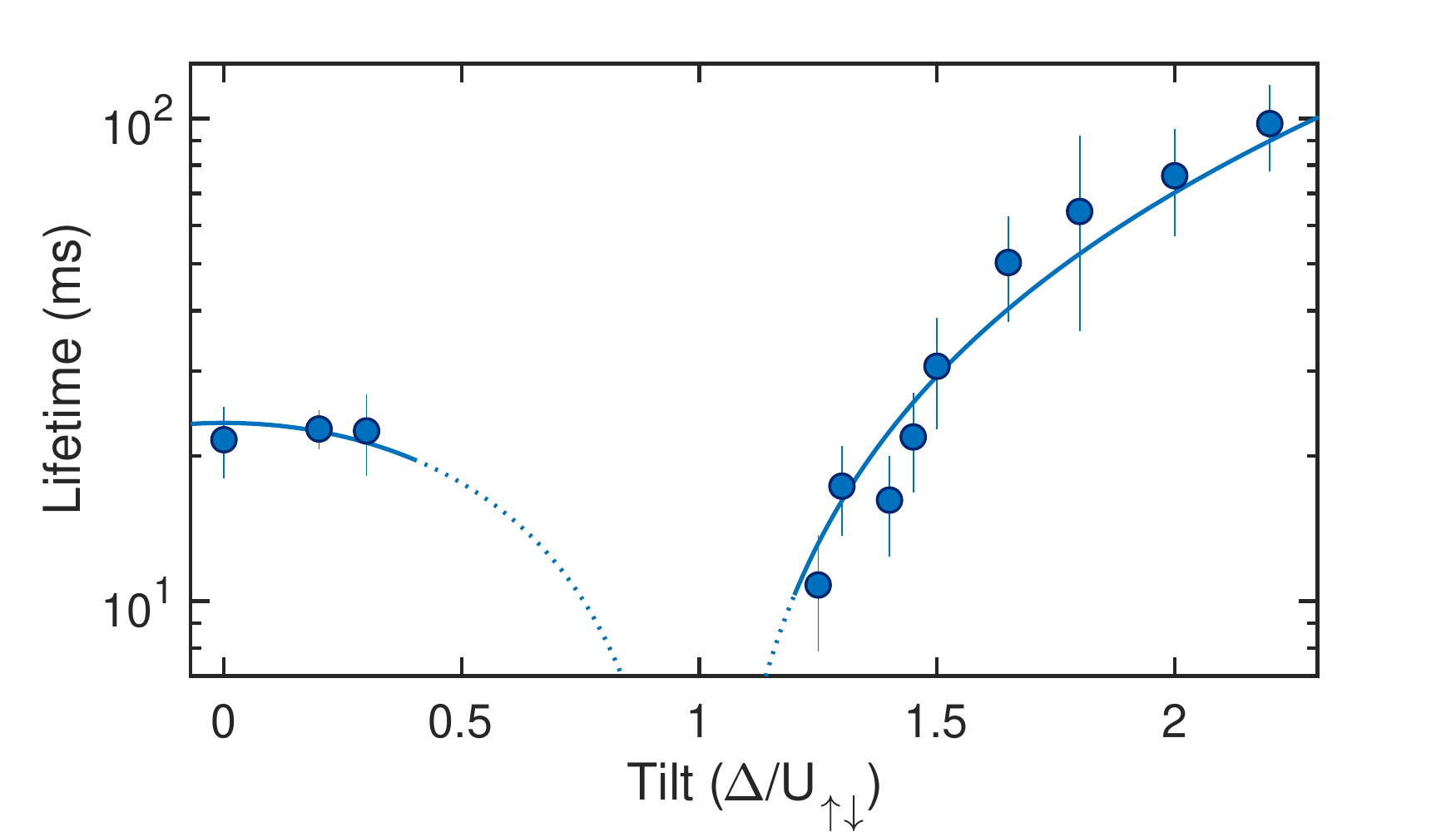}\put(0,49){\bf{b)}}
\end{overpic}
\caption{Relaxation of a non-equilibrium spin pattern by superexchange, controlled by the tilt. a) Lifetime as a function of lattice depth $V_z$ along the tilt for $\Delta\,{=}\,1.65\,U_{\uparrow\downarrow}$. Inset: decay of the contrast of the $|{\uparrow}\rangle$ state for $6\,E_R\,{\leq}\,V_z\,{\leq}\,17\,E_R$. The lifetimes are obtained from exponential fits with an offset.  b) Lifetime as a function of applied tilt at $V_z\,{=}\,12\,E_R$. The solid lines in both subfigures are $A\,\hbar\,/J_{xy}(\Delta)$, with one fit parameter: a) ${A\,{=}\,7.54\,{\pm}\,0.31}$ b) ${A\,{=}\,6.54\,{\pm}\,0.34}$. The dotted line indicates the region where the single-band approximation of the Hubbard model breaks down due to resonances of $\Delta\,{=}\,U_{\uparrow\uparrow}, U_{\uparrow\downarrow}, U_{\downarrow\downarrow}$. }
\label{spiral}
\end{figure}

A spin-1/2 Heisenberg model \cite{ddl03} is implemented using the lowest two hyperfine states of $^7$Li in a high magnetic field 
\begin{equation}
H = J_z \sum_{\langle i,j \rangle} S_i^zS_j^z + J_{xy} \sum_{\langle i,j \rangle} \left(S_i^xS_j^x + S_i^yS_j^y \right)
\label{Heisenberg}
\end{equation}
where $\langle i,j \rangle$ denote nearest neighbors, $S^\alpha_i$ are spin matrices, and $J_{z}$ and $J_{xy}$ are the superexchange parameters (see \cite{supp}). Similarly to \cite{thywissen14, munich14} (our preparation is described in \cite{supp}), we create a spin pattern and study its relaxation. Using $\pi/2$ pulses and a pulsed magnetic gradient, a spiral spin pattern is created resulting in a sinusoidal (cosinusoidal) variation of the $z$ ($x$) projection of the magnetization, which is a superposition of many spin waves (magnons), and is therefore not an eigenstate. The spiral has a pitch of 11.5$\,\mu$m, and about two periods fit within the cloud. We measure the relaxation of the spiral by imaging the decaying contrast of the real-space density distribution of $|{\uparrow}\rangle$ atoms on a CCD camera (with 4$\,\mu$m resolution) in the presence of a tilt $\Delta\,{=}\,1.65\,U$. 

We first show that the tilt does not inhibit superexchange. To simplify the interpretation, we pick a magnetic field of 848.1014\,Gauss at which $J_z\,{=}\,0$ and the dynamics are solely determined by $J_{xy}\,{=}\,J$ from Eq.\,\eqref{JDelta} with $U\,{=}\,U_{\uparrow\downarrow}$. The inset in Fig.\,\ref{spiral}a shows the decay of the contrast at several lattice depths, which collapse onto a single curve when the time is rescaled by $\hbar/J_{xy}$.  This confirms, over a range of more than two orders of magnitude ($0.015\,\rm{kHz}\,{<}\,J_{xy}/\hbar\,{<}\,2.68\,\rm{kHz}$), that the spin relaxation is driven by superexchange. We note that the contrast decays to a long-lived offset, which is larger at higher temperatures probably due to the presence of holes. Also, the offset in the spin-density modulation is smaller than the observed offset in the contrast since our imaging method enhances small contrasts. The dependence of the relaxation time and the offset on parameters of the system, such as the anisotropy of the Heisenberg model, the pitch of the spiral and temperature will be addressed in a future study.  

We now demonstrate the modification of the superexchange rate with tilt. In general, changing the strength of the tilt also changes the ratio $J_z/J_{xy}$, which determines the nature of the dynamics and the ground state. For example, when $\Delta\,{>}\,U$, the sign of the Heisenberg parameters can be flipped (see Eqs.\,S6-S7 in \cite{supp}), making it possible to go between ferromagnetic and antiferromagnetic coupling. Here we pick a magnetic field of 857.0052\,Gauss at which $U_{\uparrow\uparrow}\,{=}\,{-}\,U_{\downarrow\downarrow}$ so that $J_z/J_{xy}\,{=}\,{-}\,1$ is constant as a function of tilt. Then, varying the tilt only changes the speed of the dynamics and not the nature of the Hamiltonian. Fig.\,\ref{spiral}b shows that the relaxation times can be tuned by the tilt by an order of magnitude ($0.067\,\rm{kHz}\,{<}\,J_{xy}/\hbar\,{<}\,0.605\,\rm{kHz}$).

\begin{figure}[t!]
\begin{overpic}[width=0.5 \textwidth]{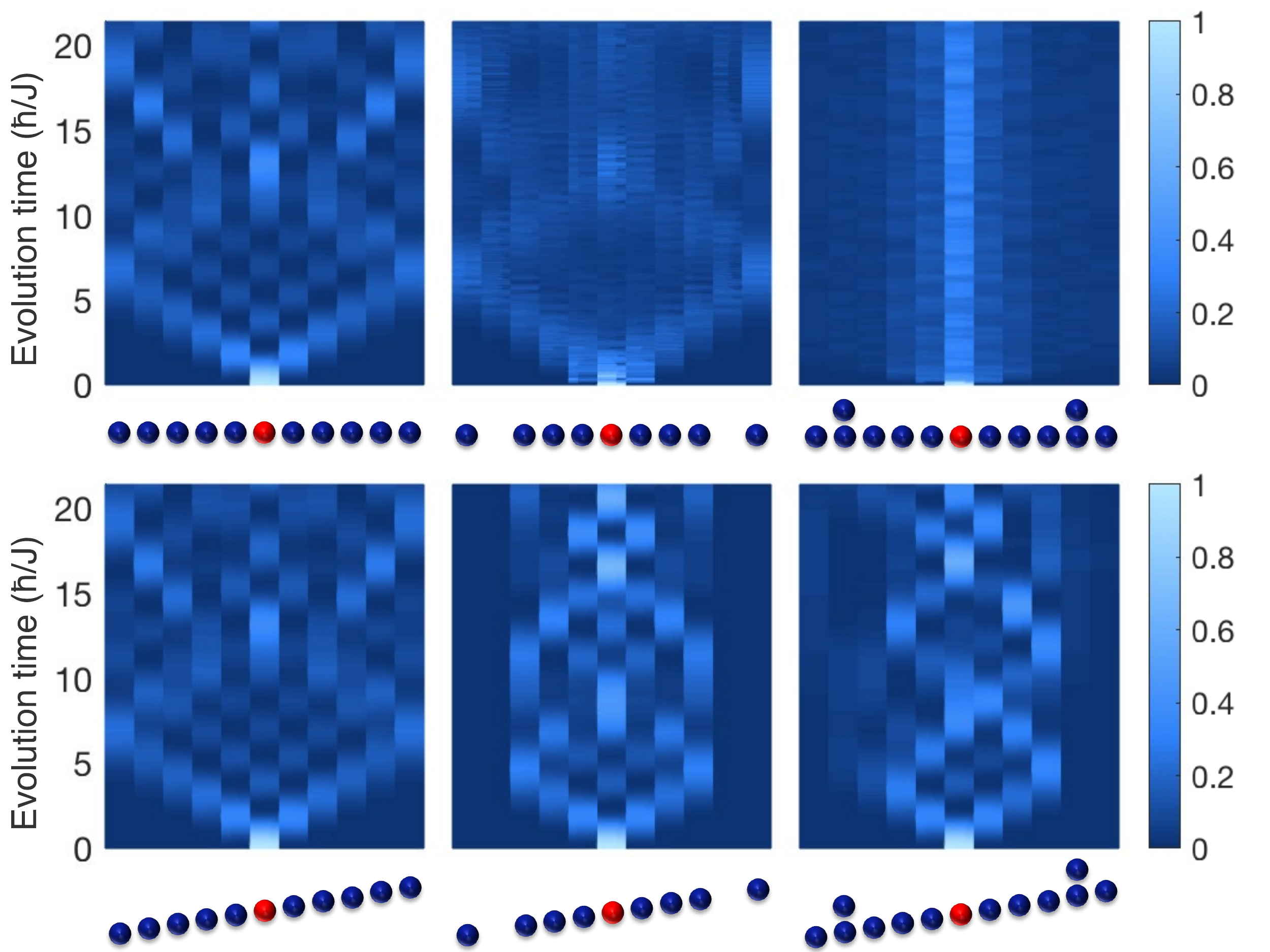}
\put(8.5,46.5){\color{white} {\bf a)}} \put(36.6,46.3){\color{white} {\bf b)}} \put(64.6,46.5){\color{white} {\bf c)}}
\put(8.5,8.9){\color{white} {\bf d)}} \put(36.6,8.9){\color{white} {\bf e)}} \put(64.6,8.9){\color{white} {\bf f)}}
\end{overpic}
\caption{Effect of holes and doublons on the superexchange dynamics of an $|{\uparrow}\rangle$ spin in a chain of $|{\downarrow}\rangle$ spins. Plotted is the probability distribution of the $|{\uparrow}\rangle$ spin as a function of time for three initial states without (top row) and with (bottom row) the tilt: no defects (left), two holes (middle), two doublons (right). Here $\Delta\,{=}\,1.25\,U$, $U$ is spin independent, and the parameters are chosen so that the superexchange rate $J/\hbar$ is the same as in the case with no tilt.}
\label{dynamics}
\end{figure}
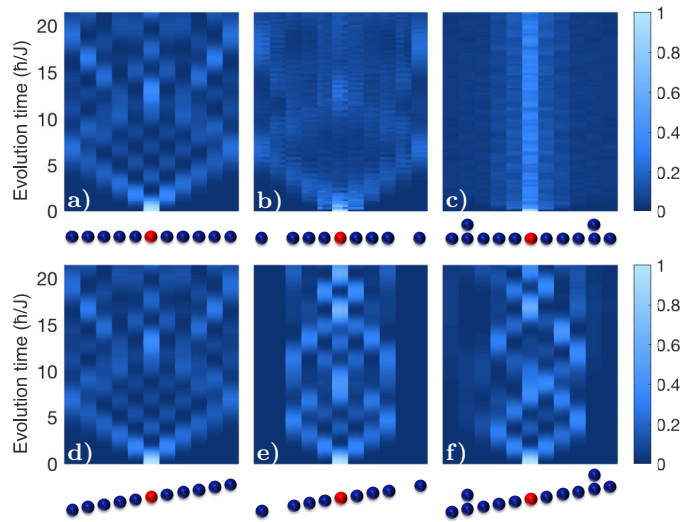

\textit{(iv) Freezing in defects.}
A direct consequence of the absence of first-order tunneling in a tilted MI is that defects, which normally propagate at a rate $\sim t/\hbar$, are frozen in. Here we illustrate the different effects of mobile and immobile holes and doublons on the spin transport of a single $|{\uparrow}\rangle$ atom in a chain of $|{\downarrow}\rangle$ atoms. We numerically simulate the evolution of the two-component Bose-Hubbard model (see \cite{supp}) for three inital states after tunneling is suddenly switched on. When there are no defects (Fig.\,\ref{dynamics}a,d), the dynamics are the same with and without the tilt. The time evolution of spin $|{\uparrow}\rangle$ shows coherent ballistic expansion of the wavefront with a characteristic checkerboard pattern \cite{fukuhara13}, akin to the dynamics of a single particle in a non-tilted lattice \cite{preiss15}. The effect of mobile holes (Fig.\,\ref{dynamics}b) is to displace the particles without impeding the overall dynamics significantly, which was also observed for antiferromagnetic chains \cite{hilker17}. Some coherent oscillations appear blurred and are restored by the tilt. In the tilted case, the holes act as domain walls, confining the dynamics to a shorter chain (Fig.\,\ref{dynamics}e). 

The effect of doublons is more subtle. Compared to holes, the presence of $|{\downarrow\downarrow}\rangle$ doublons enables the formation of an $|{\uparrow\downarrow}\rangle$ doublon, so that the $|{\uparrow}\rangle$ spin can propagate at $t/\hbar$. Note that due to Bose enhancement, an $|{\uparrow\downarrow}\rangle$ doublon quickly turns into a $|{\downarrow\downarrow}\rangle$ doublon. Fig.\,\ref{dynamics}c shows that the $|{\uparrow}\rangle$ spin is localized near the original position by collisions with $|{\downarrow\downarrow}\rangle$ doublons, which we suspect is due to destructive interference of all paths. With the tilt, doublons are pinned and act as reflective barriers. The superexchange rate for spin $|{\uparrow}\rangle$ to become part of a doublon is $J_2\,{=}\,2t^2 \left[ {-}2/\Delta\,{+}\,2/(2U_{\uparrow\downarrow}\,{+}\,\Delta) \right]$, which is different for $\pm \Delta$ and leads to the left-right asymmetry in Fig.\,\ref{dynamics}f. The effects of fixed and mobile defects in higher dimensions will be somewhat different, but overall, mobile defects can have a significant effect on spin dynamics, while immobile defects act, to a good approximation, as domain walls or static impurities. This has implications not only for dynamics, but also for adiabatic state preparation where the tilt prevents defects from increasing the final entropy (see Fig.\,S5 in \cite{supp}).

The implementation of tilts for heavier atoms, should be less demanding since similar tilts (in units of recoil energy) require lower laser power. Magnetic tilts are also possible if the two spin states have the same magnetic moment. Separation of spin and mass transport could also be achieved with random offsets implemented with bichromatic lattices or laser speckle, as in the studies of Anderson localization \cite{roati08, billy08}.

We have introduced tilted lattices as a new tool with practical and fundamental applications. On the practical side, we have shown that it can lead to an order of magnitude larger systems with spin-spin couplings which are an order of magnitude faster.  On the fundamental side, the tilt can change not only the speed of superexchange, but also the anisotropy of Heisenberg models. It also turns $t\text{-}J$ models with mobile holes into spin systems with pinned impurities. This can be used to create lattices with disorder, similar in spirit to disorder in species-dependent lattices created by pinning the second species \cite{gavish05}, and to study mixed-dimensional transport in 2D systems with tilt along one axis \cite{grust18}. The separation between spin and density dynamics should be useful for future quench experiments and for improving the fidelity of adiabatic preparation of magnetically-ordered ground states. 

\begin{acknowledgments}
We acknowledge support from the NSF through the Center for Ultracold Atoms and Award No. 1506369, from ARO-MURI Non- Equilibrium Many-Body Dynamics (Grant No. W911NF- 14-1-0003), from AFOSR-MURI Quantum Phases of Matter (Grant No. FA9550-14-1- 0035), from ONR (Grant No. N00014-17-1-2253), and from a Vannevar-Bush Faculty Fellowship. Part of this work was done at the Aspen Center for Physics, which is supported by National Science Foundation grant PHY-1607611. We acknowledge support from the EPSRC Programme Grant DesOEQ(EP/P009565/1), and by the EOARD via AFOSR grant number FA9550-18-1-0064. Results were obtained using the EPSRC funded ARCHIE-WeSt High Performance Computer (EP/K000586/1).
\end{acknowledgments}

\bibliography{gradient_bib}

\pagebreak
\newpage
\begin{center}
\textbf{\LARGE Supplemental material}
\end{center}
\setcounter{equation}{0}
\setcounter{figure}{0}
\setcounter{table}{0}
\renewcommand{\theequation}{S\arabic{equation}}
\renewcommand{\thefigure}{S\arabic{figure}}
\renewcommand{\bibnumfmt}[1]{[S#1]}

\section{Realization of tilted potentials}
The tilt is implemented by an AC Stark shift gradient across the cloud. We use a 1064$\,$nm beam, far off detuned from the 671$\,$nm D-lines of lithium so that the two hyperfine states used as spin states feel the same potential gradient. Using an optical beam makes it possible to switch the tilt on and off suddenly (within 1$\,\mu$s) with an acousto-optic modulator and to change the axis along which the tilt is applied. An alternative approach is to use a magnetic field gradient, which provides for better alignment stability and homogeneity of the tilt across the sample. However, the use of a magnetic field gradient can make it more difficult to implement large tilts, due to geometrical and power constraints (for lithium $150\,$Gauss/cm are needed for a tilt of $10\,$kHz/site), and also to make the tilts the same for all states used as pseudo-spins, due to differential magnetic moments.

 The generalization to 2D and 3D tilts is straightforward except that care has to be taken to avoid resonant second-order tunneling (see also \cite{sachdev11}). These processes arise when nearest-neighbor sites have the same potential energy offset. For example, when the tilt is along the (1,1) direction in a 2$\,$D lattice, all sites connected by a line parallel to (1,-1) have the same potential energy and a two-step tunneling process can take place, competing with superexchange. This issue can be resolved by missaligning the tilt from the (1,1) direction.     

\section{Calibration of the tilt}
\begin{figure}[t!]
\center
\includegraphics[width=0.45 \textwidth]{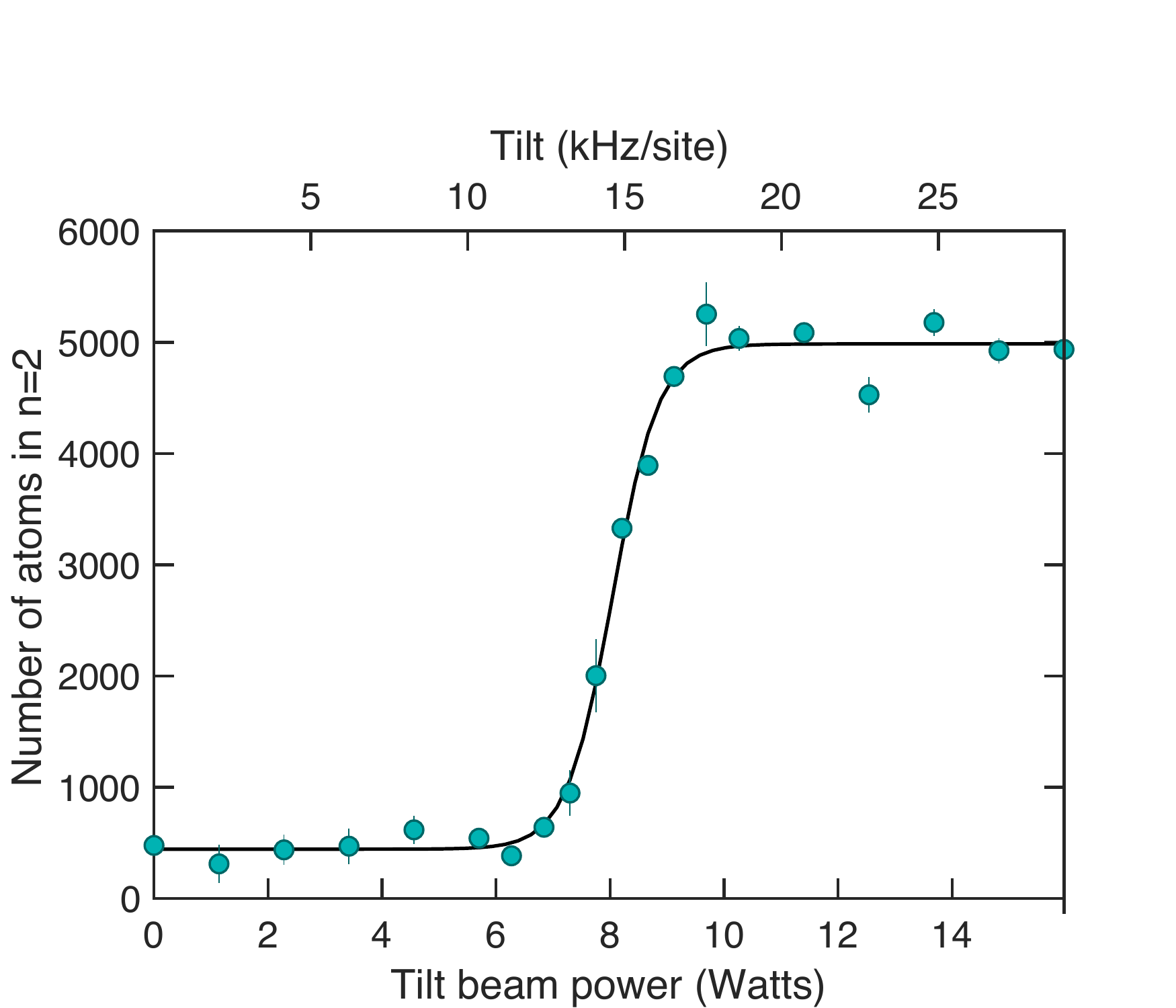}
\caption{Calibration of the tilted potential. Plotted are the number of atoms in the $n\,{=}\,2$ MI shell (with lattice depths $(V_x,V_y,V_z)\,{=}\,(35,35,12)\,E_R$) after an adiabatic ramp of the tilt as a function of the final strength of the tilt beam in Watts. After identifying the center of the transition with U, we obtain the power of the tilt in unites of kHz/site (top axis). }
\label{calibration}
\end{figure}

The tilt beam is a Gaussian beam with 73$\,\mu$m $1/e^2$ radius. After preparing an $n\,{=}\,1$ MI in a 3D lattice of 35$\,E_R$, 1D dynamics are realized by lowering the lattice along the tilt direction to 12$\,E_R$ and keeping the other two lattices at 35$\,E_R$. The calibration is done by then adiabatically ramping up the tilt across the $\Delta\,{=}\,U$ resonance. As a result, doublons are formed on every other site. This reproduces the experiments in \cite{greiner11, nagerl13}. Fig.\,\ref{calibration} shows the number of atoms in $n\,{=}\,2$ sites as a function of tilt power. We fit a phenomenological function
\begin{equation}
f(x)=A\, \tanh\left( \frac{x-x_0}{w} \right) + c
\label{fx}
\end{equation} 
 and identify the center $x_0$ of the transition with $\Delta\,{=}\,U$. For slower ramp speeds we see a second resonance at half the power, which we interpret as $\Delta\,{=}\,U/2$.\par

\section{Alignment and Inhomogeneity}
\begin{figure}[h!]
\center
\begin{overpic}[width=0.45 \textwidth]{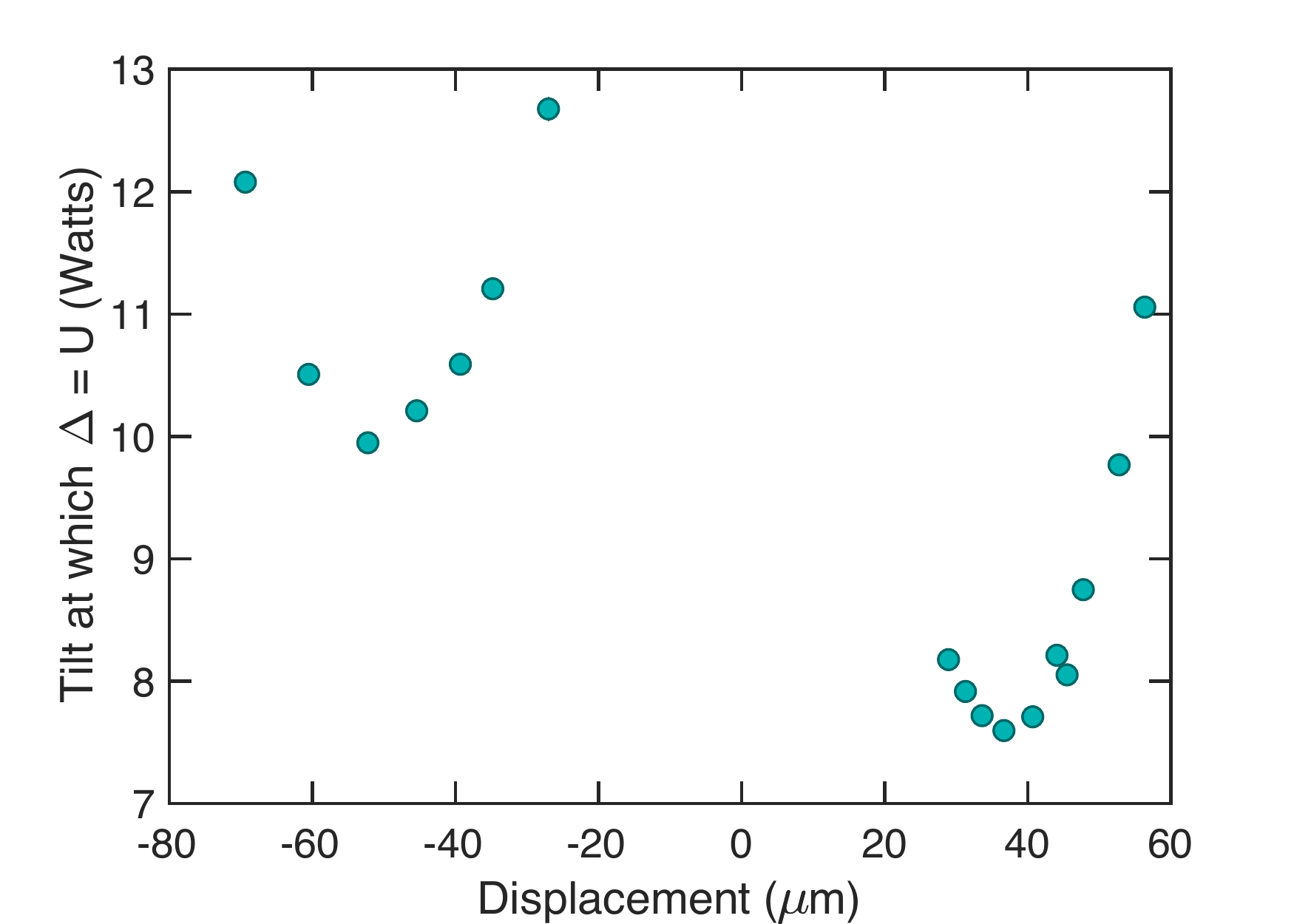}\put(-2,65){\bf{a)}}\end{overpic}
\begin{overpic}[width=0.45 \textwidth]{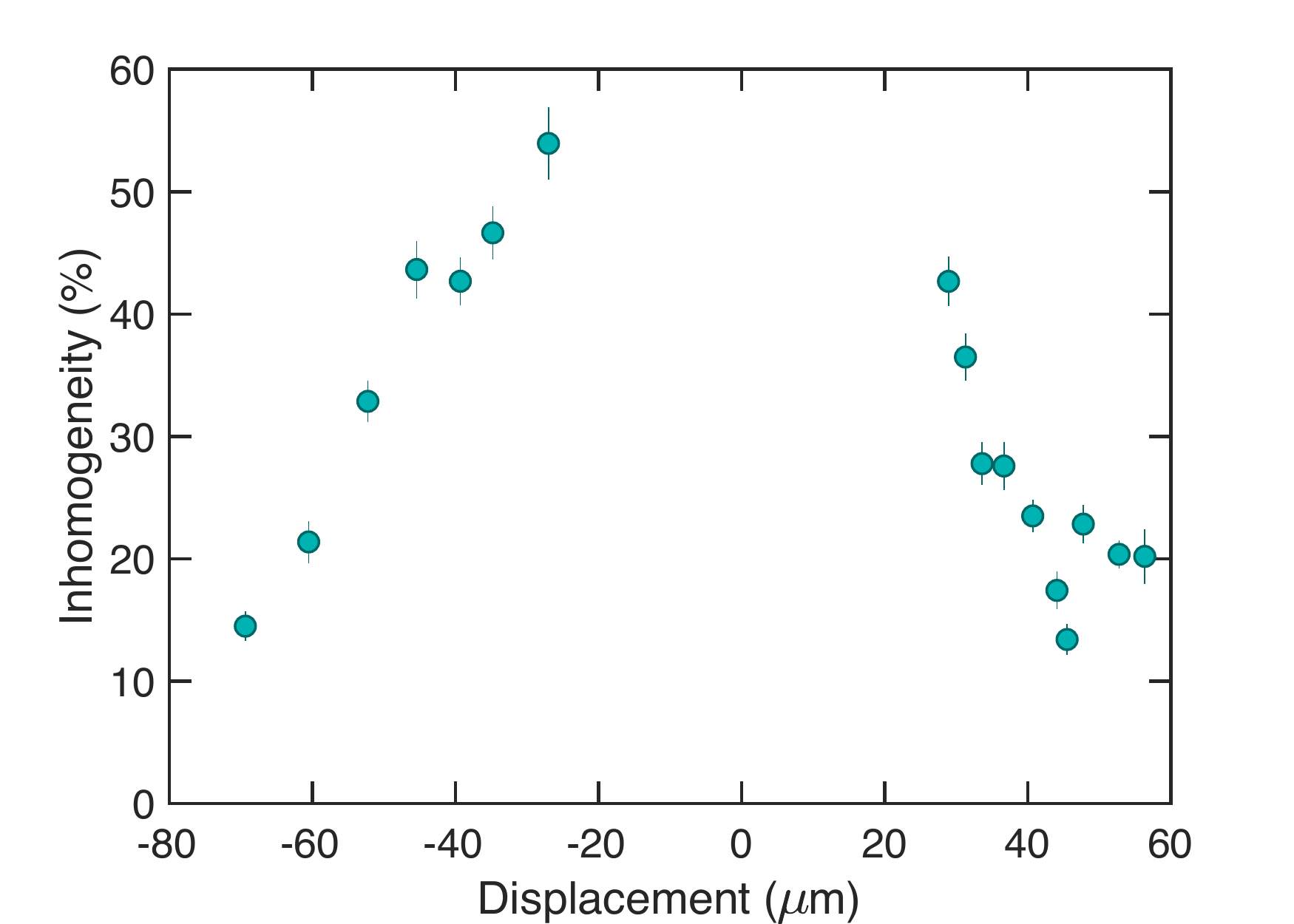}\put(-2,65){\bf{b)}}\end{overpic}
\caption{Strength and inhomogeneity of the tilt (at lattice depths $(V_x,V_y,V_z)\,{=}\,(35,35,12)\,E_R$). The extracted a) center point $x_0$ and b) width $2.2 \, w/x_0$ as a function of the displacement of the tilt beam from the atoms.}
\label{distance}
\end{figure}
We repeat the calibration measurement as a function of the distance between the MI and the center of the tilt beam. The center $x_0$ and width $2.2 \, w/x_0$ of the fitted function $f(x)$ from Eq.\,\eqref{fx} are shown in Fig.\,\ref{distance}. For a perfect Gaussian beam
\begin{equation}
g(z)=A\,e^{-2(z-z_0)^2/\sigma_0^2}
\end{equation} 
we expect that the tilt is maximized when the beam is $z_0\,{=}\,{\pm}\sigma_0/2\,{=}\,{\pm} \,36.5\,\mu$m away from the cloud. However, due to imperfections in the beam, we find it at $z_0\,{=}\,{+}\,36.3\,\mu$m and $z_0\,{=}\,{-}\,49.8\,\mu$m.  see Fig.\,\ref{distance}. 

The potential across the cloud is determined by the sum of the AC-Stark shifts of the tilt beam combined with the lattice beams. Since they are all Gaussian, their curvatures lead to an inhomogeneity of the tilt per site across the cloud. In fact, for our geometry, the majority of the inhomogeneity comes from the lattice beams, each with a $1/e^2$ radius of 125$\,\mu$m. In principle, for a given tilt beam power, it is possible to find a displacement of the tilt beam which leads to a cancellation of the curvature of the lattice and the curvature of the tilt beam at the position of the cloud. We characterize the inhomogeneity by the width of the region over which doublons form when we perform the calibration measurement in Fig.\,\ref{calibration}. We define the width as the region over which the fit function $f(x)$ goes from 10$\,\%$ to 90$\,\%$ of the asymptotic values. We note that the point of minimum width does not exactly coincide with the point of maximum tilt, as shown in Fig.\,\ref{distance}, and that for lower powers of the tilt beam, the inhomogeneity increases, which is probably due to the partial cancellation of the lattice and tilt beam curvatures. The inhomogeneity can be decreased by using larger beams, which is an option for heavier atoms for which the laser power is not so limiting as for lithium. For our experiments, we use a displacement of $z_0\,{=}\,{+}\,45\,\mu$m to minimize the inhomogeneity of the tilt. 

\section{Choice of tilt value}
For most of the experiments, we pick a tilt per site of $\Delta\,{=}\, 1.65\,U$. This choice avoids resonant tunneling and formation of doublons at $\Delta\,{=}\,U/m$ for tunneling $m$ sites away. The most prominent such resonance is at $\Delta\,{=}\,U$ and tunneling of up to 5 sites away has been observed \cite{nagerl14}. Due to the inhomogeneity of the tilt across the cloud of 10-15$\,\%$, we pick a $\Delta\,{>}\, U$ to avoid any resonances within the cloud. The scaling of the superexchange rate with tilt $x\,{=}\,\Delta/U$ is
\begin{equation}
h(x)= \frac{1}{2}\left(\frac{1}{1+x}+\frac{1}{1-x}\right)
\end{equation}
 For $x\,{>}\,1$, the sign of $h(x)$ is flipped. For $1\,{<}\,x\,{<}\,\sqrt{2}$, the magnitude of superexchange is increased. For $x\,{>}\,\sqrt{2}$ the superexchange rate decreases, for example to 50$\,\%$ at $x\,{=}\,1.75$, implying that the most useful range of applicability of the tilt is between $1\,{<}\,x\,{<}\,2$ and any points $x\,{<}\,1$ that avoid resonances.  

\section{Loading large MI plateaus}
\begin{figure}[h!]
\center
\includegraphics[width=0.45 \textwidth]{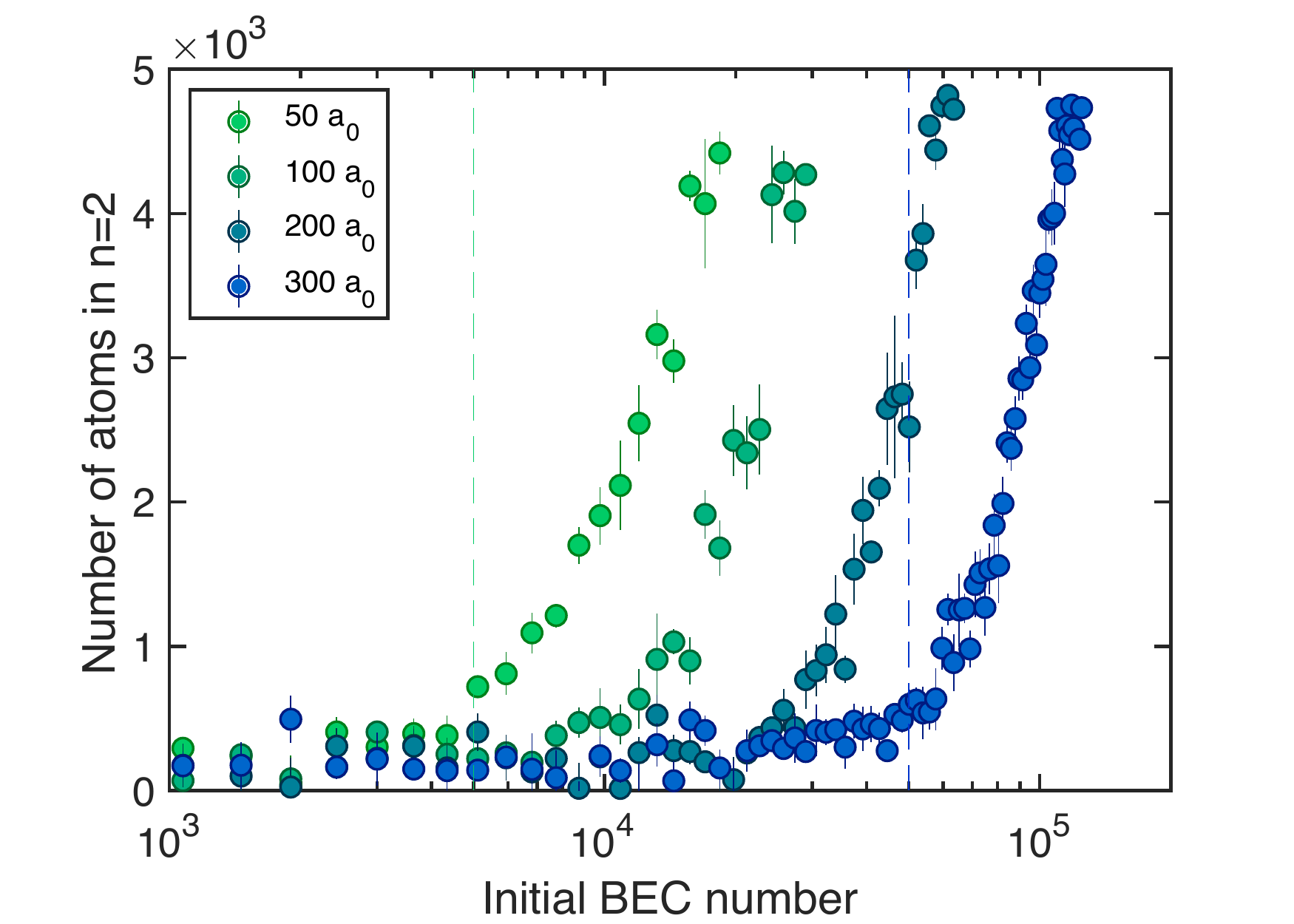}
\caption{Size of $n\,{=}\,1$ MI plateaus. Plotted is the number of atoms in the $n\,{=}\,2$ MI shell as a function of initial BEC number (with condensate fraction of more than 95$\,\%$). This is a measure of the size of the $n\,{=}\,1$ MI plateau at each scattering length given the trapping potential, which in this case is determined solely by the curvature of the lattice beams. The dashed lines indicate the maximum number of BEC atoms which fit in the $n\,{=}\,1$ MI shell at a scattering length of 50$\,a_0$ and of 300$\,a_0$.}
\label{large_numbers}
\end{figure}
To determine the maximum atom number for the $n\,{=}\,1$ MI shell, we probe the formation of the $n\,{=}\,2$ MI shell by loading successively more atoms at each scattering length and measuring the number of doubly-occupied sites: Fig.\,\ref{large_numbers}. The smaller the scattering length, the smaller the number of atoms that fill the $n\,{=}\,1$ MI plateau for a harmonic trapping potential. The maximum scattering length which can be used is the one for which the three body loss rate becomes comparable to the inverse lattice loading time.

\section{Preparation of the spin spiral}
The spin spiral is created by turning on a magnetic field gradient of 50.8$\,$Gauss/cm and then quickly applying a $\pi/2$ pulse to rotate the spins from the $|\,{\uparrow} \rangle$ to the $\left( |\,{\uparrow} \rangle \,{+}\,|\,{\downarrow} \rangle \right)/2$ state on each site. During the next 550$\,\mu$s of free evolution, the spins precess by a different amount on each site because of the magnetic field gradient and the differential magnetic moment of 31$\,$kHz/Gauss between the $|\,{\uparrow}\rangle$ and the $|\,{\downarrow} \rangle$ states at 882$\,$Gauss. Another $\pi/2$ pulse rotates the spiral into the $xz$ plane, after which the magnetic field gradient is turned off. This results in a spiral with wavelength $\lambda_{s}\,{=}\,h/(\Delta\mu B' T)\,{=}\,11.5\,\mu m$ where $T$ is the evolution time.

\section{Spiral image analysis}
 \begin{figure}[h!]
\center
\includegraphics[width=0.45 \textwidth]{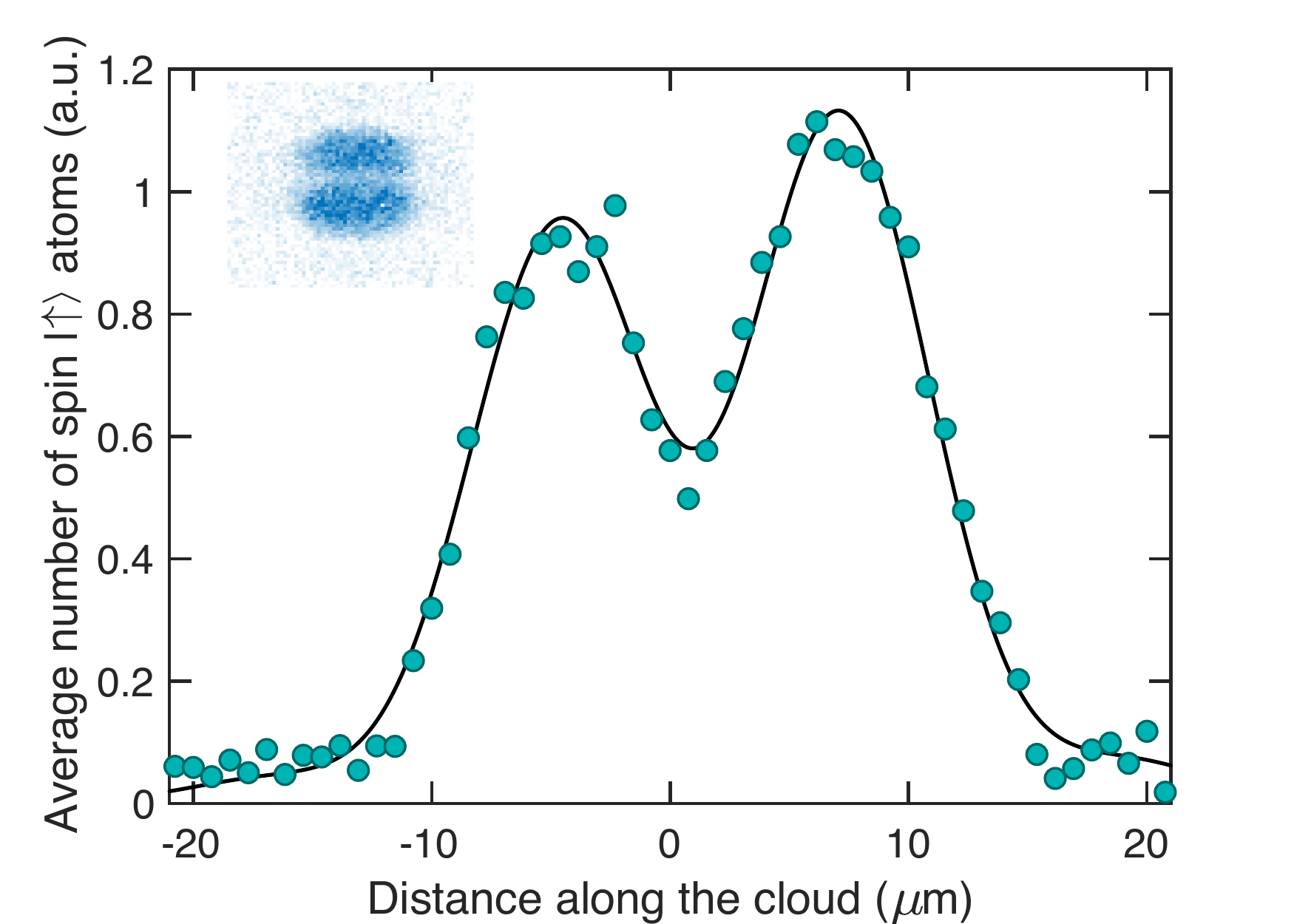}
\caption{Fitted density distribution from a single image of spin $|\,{\uparrow}\rangle$ atoms after averaging along the horizontal direction.}
\label{fit}
\end{figure}
We analyze the spin spiral by imaging one of the two spin states. Fig.\,\ref{fit} shows the spin density distribution of the $|\,{\uparrow} \rangle$ atoms. We take 5 images and after averaging along the direction perpendicular to the stripe pattern, we fit them simultaneously with a 1D function of the form:
\begin{equation}
s_j(x) = \frac{a}{2}\left[1+C\sin(Qx+\phi_j)\right]e^{-(x-x_0)^2/2w_0^2}
\end{equation}  
where only the phase $\phi_j$ is allowed to vary from image to image. Here $Q$ is the spiral wavevector, $C$ is the contrast, $a$ is an overall scaling factor, and $x_0$ and $w_0$ are the midpoint and the width of the Gaussian envelope respectively. The phase variation comes from magnetic field fluctuations which are currently $1\,\times\, 10^{-5}$ at 880 Gauss. Once we have calibrated the wavelength of the spiral as a function of evolution time, we also fix the wavevector $Q$ in the fit. We extract the contrast $C$ and rescale it to the initial contrast in Fig.\,3. The value of the starting contrast is limited by our imaging system.

\section{Heisenberg Model}
The Heisenberg Hamiltonian is derived from the two component Bose-Hubbard model with one particle per site as in \cite{ddl03}.
\begin{equation}
H = J_z \sum_{\langle i,j \rangle} S_i^zS_j^z + J_{xy} \sum_{\langle i,j \rangle} \left(S_i^xS_j^x + S_i^yS_j^y \right)
\label{Heisenberg}
\end{equation}  
where the spin matrices $S^{\alpha}_i$ are defined as $S^z_i\,{=}\,(n_{i\uparrow}\,{-}\,n_{i\downarrow)}/2$, $S^x_i\,{=}\,(a^{\dagger}_{i\uparrow}a_{i\downarrow}\,{+}\,a^{\dagger}_{i\downarrow}a_{i\uparrow})/2$, and $S^y_i\,{=}\,{-}\,i( a^{\dagger}_{i\uparrow}a_{i\downarrow}\,{-}\,a^{\dagger}_{i\downarrow}a_{i\uparrow})/2$. The coefficients are:\\
\begin{align}
J_z&=\frac{4t^2}{U_{\uparrow\downarrow}}\frac{1}{2}\left(\frac{1}{1-\Delta/U_{\uparrow\downarrow}}+\frac{1}{1+\Delta/U_{\uparrow\downarrow}} \right) \nonumber \\
&-\frac{4t^2}{U_{\uparrow\uparrow}}\frac{1}{2}\left(\frac{1}{1-\Delta/U_{\uparrow\uparrow}}+\frac{1}{1+\Delta/U_{\uparrow\uparrow}} \right) \nonumber \\
&-\frac{4t^2}{U_{\downarrow\downarrow}}\frac{1}{2}\left(\frac{1}{1-\Delta/U_{\downarrow\downarrow}}+\frac{1}{1+\Delta/U_{\downarrow\downarrow}} \right)
\end{align}
\begin{align}
J_{xy}&=-\frac{4t^2}{U_{\uparrow\downarrow}}\frac{1}{2}\left(\frac{1}{1-\Delta/U_{\uparrow\downarrow}}+\frac{1}{1+\Delta/U_{\uparrow\downarrow}} \right) 
\end{align}

\section{Numerical Simulations}
For simulating the dynamics of spin $|{\uparrow}\rangle$ in a chain of spin $|{\downarrow}\rangle$, we use the two-component Bose-Hubbard model:\\
\begin{align}
H=&-t \sum_{\langle i,j \rangle} \left(a^{\dagger}_{\uparrow ,\,i}a_{\uparrow ,\,j} + a^{\dagger}_{\downarrow ,\,i}a_{\downarrow ,\,j} \right)+U\sum_i n_{\uparrow ,\, i} n_{\downarrow ,\, i} \nonumber\\
&+\frac{U}{2}\sum_i \left( n_{\uparrow ,\, i}(n_{\uparrow ,\, i}-1)+n_{\downarrow ,\, i}(n_{\downarrow ,\, i}-1) \right) \nonumber \\
&- \Delta \sum_i i\left(n_{\uparrow ,\, i}+n_{\downarrow ,\, i} \right)
\label{Hamiltonian_delta}
\end{align}
where $a^{\dagger}_{\sigma ,\,_i}$ and $a_{\sigma ,\,i}$ are the creation and annihilation operators of spin $\sigma$ on site $i$, $t$ is the tunneling matrix element, $U$ is the on-site interaction, and $\Delta$ is the tilt. Here $U_{\uparrow\uparrow}\,{=}\,U_{\uparrow\downarrow}\,{=}\,U_{\downarrow\downarrow}\,{=}\,U$. We perform quench simulations from initial product states with and without a tilt to investigate the effects of holes and doublons on the superexchange dynamics. 

In particular, we use a one-dimensional lattice of 11 sites with three initial states: (i) unit filled MI state of $|\,{\downarrow}\rangle$ particles with a single $|\,{\uparrow}\rangle$ particle on site 6; (ii) we replace the $|\,{\downarrow}\rangle$ particles on sites 2 and 10 with holes; (iii) we replace holes with $|\,{\downarrow\downarrow}\rangle$ doublons. We time-evolve the initial state either with no tilt $\Delta\,{=}\,0$ or with a tilt tuned close to the resonance $\Delta\,{=}\,1.25$. Calculations were performed using the open-source Python package for exact diagonalization and quantum dynamics QuSpin v0.3.2 \cite{marin17, marin18}.

\section{Adiabatic state preparation with a tilt}

One of the important potential applications of preventing transport of holes and doublons is in isolating entropy, especially when the entropy is initially dominant at the outside of a harmonic trap. For example, if we begin with a Mott Insulator plateau in the center of the system, we can adiabatically prepare interesting spin-ordered states within that plateau \cite{sorensen10, schachenmayer15}. However, this situation is complicated if propagation of entropy into the Mott Insulator plateau (as holes and doublons) is allowed during the adiabatic ramp. In the presence of a tilt, these become pinned defects, as described in the main text. This will generally isolate entropy at the edge of the system, and in the worst case it will generally lead to a break for holes, or weak coupling for doublons in the spin chain.

A full analysis of this situation for a general thermal state would be an interesting next direction. As a simple demonstration, here we consider adiabatic state preparation in the presence of a single hole, with and without a tilt. We aim to begin with all spins aligned with the x-axis, and then prepare an XY Ferromagnet within the Hamiltonian of Eq.\,\eqref{Heisenberg} \cite{altman03}. This can be accomplished by switching on an RF field that couples the spins, initially far detuned from resonance, and then tuning the field into resonance. The difficult part of the adiabatic process is the removal of this field, which is what we model here. The modified Bose-Hubbard Hamiltonian is  
\begin{equation}
\label{eq:XYFM_Hamiltonian_ramp}
H_{a}(\tau) = H - \Omega(\tau) \sum_i S^x_i,
\end{equation}
where $H$ is the Heisenberg Hamiltonian (\ref{Heisenberg}) and $\Omega(\tau)$ is the time-dependent external field, which we decrease during the adiabatic ramp in the following way
\begin{equation}
\frac{{\rm d} \Omega(\tau)}{{\rm d} \tau} \sim \Delta E(\Omega(\tau)),
\end{equation}
where $\Delta E$ is the energy gap of the Hamiltonian \eqref{eq:XYFM_Hamiltonian_ramp} computed numerically.

\begin{figure}[h!]
\begin{centering}
\includegraphics[width=1\linewidth]{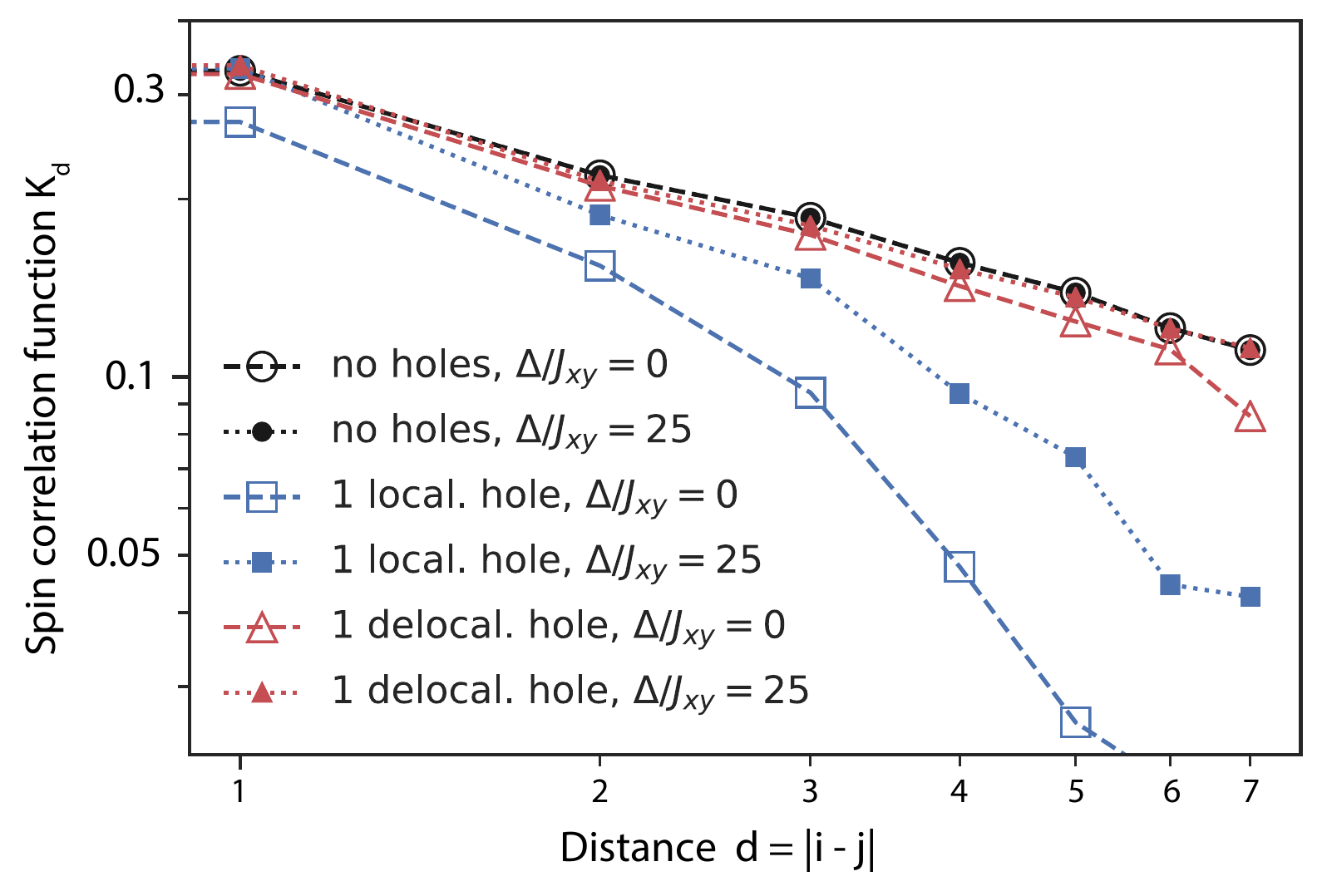}
\par\end{centering}
\caption{Averaged over distance correlation function $K_d\,{=}\, \langle |K_{i,i+d}| \rangle_i$ of the adiabatically prepared state of 10 sites with the following parameters: $J_z/J_{xy}\,{=}\, 1/5$, $t/J_{xy}\,{ =}\, 5$, $\Omega(0)/J_{xy} \,{=}\, 50$, $\tau_{\max} J_{xy} \,{=}\, 500$. The results are presented for 3 configurations of the initial state with and without tilt $\Delta$. 
\label{fig:asp_correlations}}
\end{figure}

In Fig.\,\ref{fig:asp_correlations}, we present example results, beginning with spins aligned along the x-axis, and either no holes or a single hole, which can either begin in the center of the system (localized), or begin in a superposition of $L$ states where the hole occupies different sites with equal amplitude (delocalized). In order to account for holes we compute the adiabatic ramp with the tunneling terms as well as the linear tilt terms
\begin{align}
\label{eq:XYFM_Hamiltonian_ramp_with_holes}
{\tilde H}_{a}(\tau) = & H_{a}(\tau) 
- t \sum_{\langle i,j \rangle} (a_{\uparrow,i}^\dagger a_{\uparrow,j} + a_{\downarrow,i}^\dagger a_{\downarrow,j}) \nonumber \\
&- \Delta \sum_i i (n_{\uparrow,i} + n_{\downarrow,i}),
\end{align}
where the notation is the same as in Eq.\,\eqref{Hamiltonian_delta}.

When the hole is pinned by the tilt, it creates spin chains of shorter length. As an indicator of the effects just on the adiabatic state preparation, we consider correlation functions at each possible separation distance where we remove any contributing state where a hole has broken the chain. This gives a renormalized conditional correlation matrix
\begin{equation}
K_{i,j} = \frac{C_{i,j}}{N_{i,j}},
\end{equation}
which neglects the contributions from holes. Here 
\begin{equation}
N_{i,j} = \sum_k |c_k|^2 F(i,j,k)
\end{equation}
is the renormalization matrix for $| \psi \rangle = \sum_k c_k | k \rangle$ expanded in the Fock basis $\{| k \rangle\}$ and the numerator
\begin{equation}
C_{i,j} = \sum_{k,k'} c_{k'}^* c_k F(i,j,k) \langle k' | S_i^+ S_j^- | k \rangle
\end{equation}
is the conditional correlation matrix. For both functions the condition is defined as
\begin{equation}
F(i,j,k)=1, \,\,\, {\rm iff} \,\,\, \sum_{l=i}^j \langle k | n_l | k \rangle = |i-j|+1,
\end{equation}
which takes into account only states $| k \rangle $ that do not have holes between sites $i$ and $j$.

As shown in Fig.\,\ref{fig:asp_correlations}, in the absence of holes, the tilt does not have an effect on the correlations and they decay following the expected algebraic law. When holes are added, the correlations are suppressed without a tilt, more prominently so when the hole is localized. This effect of mobile holes becomes significantly weaker in the presence of the tilt which pins the holes, restoring the strength of the correlations. For a localized hole (in the middle of the chain), the tilt effectively splits up the chain in two parts and prevents the build-up of correlations between them, which leads to suppression of correlations at distances longer than half the chain.

\end{document}